# Learning physics in context: a study of student learning about electricity and magnetism

This paper re-centres the discussion of student learning in physics to focus on context. In order to do so, a theoretically-motivated understanding of context is developed. Given a well-defined notion of context, data from a novel university class in electricity and magnetism are analyzed to demonstrate the central and inextricable role of context in student learning. This work sits within a broader effort to create and analyze environments which support student learning in the sciences.

# INTRODUCTION:

As I began teaching in a large scale, well-recognized physics department, I was struck by the fact that advanced level undergraduate physics majors, even graduate students, arrived at my class with a spotty understanding of basic concepts in electricity and magnetism. Many of these students, who had passed the introductory courses with high grades, failed to answer more than half of the questions correctly on a basic conceptual survey of the field (Finkelstein 2003). The same survey issued to students taking the introductory level course yielded results that were yet worse.[1] The most basic concepts covered in the course were not reaching the students.

Of course, this situation is not new and has been reported extensively within the physics community (Aarons 1976, Hestenes et al.1992, Hake 1998, McDermott and Redish 1999, Redish 2003) and well as by cognitive psychologists (see Reiner et al, 2000 for a review). Traditionally taught physics classes fail to impart robust conceptual understanding, even for those students who perform well on class exams. In essence, if we are concerned with developing students' conceptual understanding of physics, these classes are producing far too many false-positives. For example, faculty researchers have found that students can calculate the potential difference in a complicated, abstract circuit diagram, but can not correctly predict what happens to current, brightness, voltage, and power in a far simpler circuit using more realistic iconography (Mazur 1997). Findings such as this have spurred researchers to develop new curricula (McDermott and Schaffer 1998), rearrange course structure (Laws 1997, Goldberg and Bendall 1995), and study student learning (diSessa 1988, Redish 1994).

To date, discussion of university student learning in physics been largely student- and content-centred (McDermott and Redish1999; Redish 2003). A common goal of these efforts is to design activities that promote conceptual change in students who fail in traditional forms of instruction. While researchers go to great lengths to create environments supportive of such conceptual change,[2] the environments that promote student learning remain under-theorized. Local culture and context remain implied or alluded to in these student- and content-centred models of student learning. While the theorists of physics learning in these traditions have moved toward a more differentiated view of the learner (few would defend the view that students themselves are homogenous), these same theorists continue to treat context generically, in an undifferentiated manner, and thereby fail to provide an explicit account of how context and the teaching/learning process are interconnected. In order to understand which elements of learning environments shape students' understanding and how this is achieved, it is necessary to develop a conceptual understanding of these contexts as they interact with student learning. Consequently, this paper strives to articulate an educationally useful conception of context, and begins to develop a theory of how context participates in the process of student learning in physics.

---

[1] In a premedical introductory level physics course offered the same semester, students scored 25% correct on a multiple-choice survey of basic ideas in electricity and magnetism (E&M) given at the beginning of term. They scored 36% correct after instruction.

Socio-cultural researchers of student learning in anthropology, education, and psychology have a long-standing tradition of emphasizing the social, cultural, and context-bound nature of learning. The present work emerges from this tradition and develops a notion of context in physics learning by building particularly upon research that emphasizes the situated and embedded nature of learning (Lave 1988; Rogoff 1992), and the role of local culture and the use of cultural tools that mediate human action (Cole 1996; Engestrom 1993). This work holds context central to student learning, not as an analytically separate factor, nor as the backdrop against which student learning occurs, but as an integral part of student learning. Students (and other educational participants) shape and are shaped by the context in which these educational endeavours occur. This paper presents a model of context for examining physics education, and presents motivation for and application of this model to an environment that is specially designed to blend activities so that students engage with physics content in a manner that leads to conceptual shifts in understanding of electricity and magnetism.

## INCLUDING CONTEXT IN PHYSICS

---

[2] Reviews of physics education reforms can be found in McDermott and Redish 1999 and Redish 2003.

In recent years within the physics education research community, context has begun to emerge as topic of discussion. However, with rare exceptions, invocation of context arrives at the end of articles, as part of the agenda for the future, and its meaning remains underspecified. For example, in an important article that brings physics education and cognitive science research perspectives to the physics community, Redish concludes:

> The typical university course is a complex structure. It involves physics content, a teacher, perhaps graders or teaching assistants, a classroom, a laboratory, and, for each class, a particular set of students. Above all, it involves expectations and contexts for both the teacher and the students. If we are to make serious progress in reaching a larger fraction of our students, we will have to shift our emphasis from the physics content we enjoy and love so well to the students themselves and their learning. We must ask not only what do we want them to learn, but what do they know when they come in and how do they interact with and respond to the learning environment and content we provide (Redish 1994, p. 803).

In later work, Redish revisits these same ideas and explicitly lists context as one of his five principles: 'the context principle: What people construct [for mental models] depends on the context – including their mental state'[3] (Redish 1999, p. 565). In an example of this approach in action, Bao and Redish come back to their association of context and mental states in developing a quantitative model of student learning and performance on multiple choice tests. They state that 'the basic idea of our method is to consider that a student's knowledge is organized into productive, context-dependent patterns of association we refer to as *schemas*' (Bao and Redish 2001, p. 46). In this treatment, context remains elusive and under-specified. Context appears as a tautology – all those (external) factors that affect student schemas. Furthermore, while useful to foreground the student (and background context) in creating a standardized assessment or in evaluating conceptual models through the use of such assessment instruments, there is no guarantee that student responses will not shift given a shift in context (with any variety of definitions of context) (McCullough 2001).

---

[3] Another of Redish's principles relevant to this discussion is his 'social learning principle: For most individuals, learning is most effectively carried out via social interactions' (Redish 1999). Here he alludes to Russian psychologist, L.S.Vygostsky. Such a principle is critically important in the creation of educational environments, and, in fact, may be made more strongly: *all* learning occurs through social interactions. Vygotsky states, '*human learning presupposes a specific social nature and a process by which children grow into the intellectual life of those around them*' (Vygotsky 1978 italics in the original).

diSessa, Elby and Hammer make use of context in a different manner, making it a focal point of their work on student epistemology (diSessa et al. 2003). In a detailed case study, they demonstrate the utility of a context-dependent model of epistemology over the more traditional categorical, consistent and systemized model of student epistemology.

> What we question are theoretical frameworks and attendant methodologies that *presume* such coherence and systematicity (or choose not to examine them explicitly), and as a result may overlook evidence in students' behavior of the context-sensitive activation of finer-grained knowledge elements (diSessa et al. 2003, p. 238)

After detailing a variety of circumstances and 'epistemologically loaded behaviours' in which the student 'J' observably shifts in her beliefs, they conclude:

> We frequently found that 'beliefs' (explicit or implicit) seemed unable to carry the burden of explaining J's behaviors. Instead, we offered the idea of 'judgment in context,' which provides a first-pass at a replacement for the idea of 'exercising beliefs' with respect to what is happening on occasions when epistemological knowledge becomes active. Judgment in context serves to label why J might make one epistemological move in one context (say, split a concept), and another move in another context (say, remain accountable to a core meaning across different exemplars of a concept) (diSessa et al. 2003, p. 284).

Clearly context is critical in the epistemological development of student J. A detailed understanding of context, and the roles that it serves to affect J's behaviours and beliefs remain to be investigated.

In a related vein, Leander and Brown focus on the dynamics of student interaction, and create a framework for identifying and analyzing these student interactions in a physics classroom (Leander and Brown 1999). Their analysis examines stabilities (or lack thereof) of student dynamics and discourse from a largely activity-theoretic perspective, and hence focus on the activity-system as a fundamental unit of analysis (Engeström 1993). Their analysis demonstrates the constraints and affordances of social, institutional, classroom, and conceptual environments on the students' interaction. Each of these levels of context shapes (and is shaped by) student dynamics. While the emphasis of Leander and Brown remains on students and the stability of their interactions, by documenting the utility of their framework for studying student dynamics they implicitly invoke a differentiated notion of context. The present work strives to foreground context, distinguishing those features which help shape and are shaped by student interaction.

The approaches of Redish, of diSessa, Elby and Hammer, and of Leander and Brown provide a good point of departure for the present attempt to elaborate on the concept of context. Their discussions of contextual dependence of student understandings (of formation of knowledge schemas, epistemologies, or student dynamics) make the case for the role for context in the teaching/learning process. While these discussions emphasize students, the present work strives to make a more detailed examination of what is meant by context and how it shapes and is shaped by student learning.

**A MODEL OF CONTEXT**

At this point, it will be helpful to pause to consider exactly what is meant by the term "context," which has been described as 'devilishly polysemic' (Cole 1996, p. 338). This following examination of context follows from Cole, a cultural psychologist studying education (1996). As found in many dictionaries, context most commonly refers the surrounding situation or environment of some object or activity (Merriam-Webster 2003). Despite utility of such a definition, Ray McDermott, an anthropologist who studies classroom learning, is critical of this use of context for illuminating the teaching/learning process. McDermott uses the analogy of a bowl of soup to describe this sense of the word:

       In all common sense uses of the term, context refers to an empty slot, a container

into which other things are placed.  It is … the bowl that contains the soup. As such it [context] shapes the contours of its contents; it has its effects only at the borders of the phenomenon under analysis ... all can be analytically separated and studied on their own (R. McDermott 1993, p. 282).

McDermott criticizes this notion of context because it naturally leads to a transmissionist (source-receiver) view of the teaching/learning process and the view of the student as an empty vessel: 'By this account, knowledge and skill enter heads, where they wait passively for situations in where they might prove useful' (R. McDermott 1993, p. 282).  Nonetheless, the notion of 'that which surrounds' will prove useful, if limiting, in locating relevant scales of interaction in student learning physics.

Another sense of context is useful to consider at this point.  Cole points toward the Latin root of the term, "*contexere*, which means 'to weave together.'"(Cole 1996, p. 135)  Context arises in the weaving together of constituent elements.  Context is the collection of components and the relations among them – the connected whole which includes constituent elements *and* the relations among them.  Birdwhistell uses the analogy of a rope to develop such a notion of context:

The fibers that make up the rope are discontinuous; when you twist them together

you don't make them continuous, you make the thread continuous .... even though it

may look in a thread as though each of those particles[fibres] are going all through

it, that isn't the case... Obviously, I am not talking about the environment.  I am not

talking about inside and outside.  I am talking about the conditions of the system

(Birdwhistell as quoted in R. McDermott 1993, p. 274).

The rope is not that which surrounds the fibres, rather it is the collection of fibres and relations of the fibres with each other -- the conditions of the system.  Removing all the fibres from the rope and examining the fibres and the rope separately is not possible.  So too with this interpretation of context – the task and its context are *mutually constitutive*.  By analogy, consider the utility of 'index of refraction' which is nothing more than the collective 'weaving together' of the polarizability of individual atomic dipoles.  On the basis of a single atom, the gross or aggregate behaviour does not exist, and yet, the bending of light occurs as a result of the collection (and interaction) of atoms.[4] Here, individual learning and context a mutually constituting. Not only is conceptual ecology of the individual tied to social and physical contexts in which the individual is located, but *vice versa,* that the individual and the context create one another (Rogoff 1990, Rogoff 1992, Cole 1996). Hence, in educational environments, student learning and its context shape each other; neither may exist without the other.[5]

In foregrounding context in the discussion of student learning, this piece argues for a complementary model of context (that which surrounds and weaving together).[6] Each conception of context provides a useful lens with which to view the role of context in student learning. The container metaphor of context is particularly useful in identifying varying levels, or *frames of context*; however, without the inclusion of a dynamic, relational notion of weaving together, the notion of context remains static, analytically separable (and hence limited).

Following Cole's notions of context as containing, it is possible to adapt his "concentric circles representing the different 'levels of context' "(Cole 1996) to learning physics. Student learning in physics may be described as concentric levels of context, as shown in figure 1. For instance, a third year undergraduate may be learning about capacitance in a physics class designed for premedical students – the class contains the classroom, instructor, lectures, problem sets, tests, etc. At the centre of the figure, the student engages in a task (solving a particular problem on capacitance) which is designed to facilitate student understanding of a concept (how capacitance adds in series or parallel). The student, task (and even concept) exist within some broader context (e.g., doing problem sets), which is part of a class, which is embedded in a university with specific requirements, and so forth. It may also be observed that this representation is generally hierarchical. The outer levels more strongly direct the inner levels than *vice versa*. The course structure, pace, and audience constrain the homework sets more than the homework sets shape the class. Yet, no level exists autonomously. Each level shapes and is shaped by the levels (directly) above and below (inside and out), and thus, to study and locate a given context, one must really examine at least three levels of context (Cole 1996). For a given situation, such as working out a problem set for a given chapter, it should be recognized that the situation is part of a larger class (say, electricity and magnetism for biology majors) and that it includes certain tasks such as manipulating equations individually or in groups. Figure 1 represents only a fixed slice of embedded contexts; while the outermost level is itself embedded (in a geographic location, political and economic context), so too does the innermost level surround other levels of context.

---

[4]    aggregate behavior is what Redish refers to as a collective variable (e.g. temperature) (Redish 2003).

[5]    The ambiguity of context is what give rise to the old physics joke: upon not having seen each other for many years, one physicist says to another 'What's new?' The other replies, 'frequency.'

[6]    Physics has a fine tradition of developing multiple models of understanding of single physical properties. Most famous is the particle-wave duality of light. This dual-natured description of content is not inherently contradictory, but describes different features depending upon context.

[Insert figure 1 about here]

Using the notion of concentric or nested levels of context, below, three frames of context are used to examine student learning in physics: the particular form a task takes (as in, the context of a particular problem – whether the problem involves turntables or springs), the situation in which such action takes place, and the broader context that creates the circumstances for the particular situation. These frames of context are identified as *task*, *situation*, and *idioculture*, respectively. Reconsidering the notion of context as 'that which surrounds' results in figure 2. Tasks are embedded in situations which are located in idiocultures.

[Insert figure 2 about here]

What is identified as *task*, Mestre often refers to as context (of a problem) (Mestre 2002). He suggests that context is the story line of a problem. Following the work of Chi et al (1981) in distinguishing between expert and novice problem solving, Mestre describes the difference between the features upon which individuals focus, namely, surface features and deep structural features. Two problems that appear similar to a novice, because they both deal with sticks and clay, are not similar to the expert, who views one problem as an energy problem and the other as a force problem. Interestingly, these problems take on different task formations, or different contexts, depending upon the subject and the situation in which the problem is posed. The novice observes the problem in the context of the story line. The expert observes the problem in the context of force, momentum, and energy relations. These concepts, problems and descriptions are dynamic, woven together differently to yield differing contexts for the novice and the expert.

Mestre raises additional noteworthy points. First, his discussion of the notion of learning as transfer (of conceptual understanding) from one context to another resonates with many others (L. McDermott 1993, Engeström 1997). Following these threads, the idea of transfer across context is not limited to shifting between tasks, but instead may be applied to all frames of context (and particularly those discussed further below: situation and idioculture). Mestre himself alludes to broader notions of context. He refers to mathematics and physics contexts (which are at a scale larger than that of idioculture). Furthermore, his work focuses on the utility of having students solving problems in a new vein, where students are asked to pose problems rather than solve them for a particular class. This frame of context is identified as a situation.

The next level of context, *situation*, is borrowed from Dewey. In his forward-thinking essays of the 1930's Dewey writes:

> What is designated by the word 'situation' is *not* a single object or event or set of objects and events. For we never experience nor form judgments about objects and events in isolation but only in connection with a contextual whole. This latter is what is called a 'situation' (Dewey 1938).

And despite psychologists' efforts to reduce the study of objects or events in isolated or abstracted fashion, Dewey argues:

> In actual experience, there is never any such isolated singular object or event; *an* object or event is always a special part, phase, or aspect, of an environing experienced world – a situation . . . There is always a field in which observation of *this* or *that* object or event occurs (Dewey 1938).

Situation is the context of a task or action. It includes the participant, the task at hand, the goals and concepts of this task, the local environment and associated tools. A situation is where and how a task such as problem solving occurs. Two students working in collaboration to develop a problem on angular momentum for a particular class assignment, sitting in a classroom full of resources (a teacher, books, turntables, etc.) on a particular day constitutes a situation.

Extending the field of view allows for the collection of many situations over time that are localized to a particular group of individuals. One way of describing this frame of context is by thinking of small-group cultures, or what Fine calls *idiocultures* (Fine 1987). Fine uses this level of resolution to examine Little League Baseball, but it also appropriately describes what occurs at the level of a university course or major. Fine writes,

> Every group has its own lore or culture, which I term its idioculture... Idioculture
> consists of a system of knowledge, beliefs, behaviours, and customs shared by
> members of an interacting group to which members can refer and that serve as the
> basis for further interaction (Fine 1987).

Each course develops its own idioculture – localized customs and behaviours. Whether students expect a quiz on Fridays, or that music plays at the beginning of each physics lecture, or that every few weeks each student must report on a reading, localized meaning is constructed out of various tools and interactions. In some classes, music stopping means that class is about to begin; whereas, in others, music stopping may mean that class is over. The idioculture is the medium in which situations are embedded (and constituting). If a student is asked to pose a problem which demonstrates various physical principles, it happens in an idioculture with certain rules and norms of behaviour. In a premedical physics course, students engaged in problem posing or problem solving act within certain localized guidelines specific to that idioculture. A valid student problem generally does not come in the form of 'improve on the design of the space shuttle', nor, 'complete the following statement of Newton's Second Law: F= ____.' A task or problem that is too difficult or too simple is not considered valid because it violates the norms of the idioculture.

At the same time, the notion of context as that which surrounds is not sufficient to capture the complex and dynamic sets of relations among tasks, situations, and idioculture. Idioculture both encompasses and is constituted by the relations among situations and tasks, through their collective weaving together. The norms and collective practices in an idioculture evolve over time and are constituted by the situations that it encompasses (while at the same time, these situations are shaped by the surrounding idioculture). Furthermore, the boundaries of context and content are dynamic, shifting with goals, participants, and setting. See figure 3 below. A multitude of situations are included within a given idioculture. An idioculture spans the collection of many situations over some time span. Furthermore, situations themselves may overlap. As a result, it is possible that the object of study (content) may actually form the context of study in another situation. For example, properties of capacitors may either serve as the means or object of study in some problem.[7] . Considering the idioculture as a weaving together of situations is fruitful in order to capture the relational sense of situations and idioculture – they mutually shape one another. Attempting to remove situations completely from an idioculture is just as fruitful as extracting the fibres from the rope. One may yield a closer examination of the constituent elements (of the idioculture), but without the broader context, or weaving together of situations, they no longer have meaning and the idioculture (or rope) no longer exists.

[Insert figure 3 about here]

With this model of context, it is possible to turn to the task of moving context explicitly into the field of view when considering student learning in physics. Such an approach stems from a core principle of *contextual* constructivism:[8] *It is not fruitful to separate student learning from the context in which it occurs; context is not simply a backdrop for student learning. Rather, context is intrinsic to student learning, it shapes and is in turn shaped by both the content and student (who*

---

[7] Calculating the charge held by a given plate depends upon the context of a given capacitor geometry; meanwhile, studying the relationship between geometry and the definition of capacitance may also be an object of study.

[8] In the 1990's a variety of efforts expanded Piaget's notion of constructivism (Scott et al. 1992, Coburn 1993).

*is active in developing an understanding of a given domain and arrives with prior history).* Here it is advocated that student learning in physics is *always* intertwined with and constituted in context and inherent in a given context are certain features that promote or inhibit construction of content understanding.

## METHOD OF STUDY:

The study examines a novel class on teaching and learning physics which is offered to upper division physics majors and graduate students in physics at a large-scale research university. The ten week course was created as a form of design experiment, simultaneously used to study student learning in context and effect change (A. Brown 1992, Kelly 2003). In the present study, context is brought into the foreground in the analysis of student learning. Initially, the broad or macro-scale relations between student learning and idioculture are introduced. Following, the analysis examines three situations from a contextual constructivist perspective to focus on the interplay between student conceptual development and particular situations within which the students are engaged. While the analysis focuses on situations, these situations themselves are heavily influenced by both the idioculture and the particular task formations. Rather than simply emphasize a single unit of analysis, various forms of data are collected and analyzed at multiple tiers, or frames of context.

### I. MATERIALS:

The data collected for this study emphasize student performance, and in particular evaluate student understanding of physics (broadly conceived) and notions of teaching/ learning. Evaluation of student performance includes: pre- and post-tests of basic concepts in electricity and magnetism, audio-taped recordings of all class sessions, students' written evaluations of the course (collected fifth and tenth weeks of class), students' written "statements of teaching" (collected during first and last weeks of class), students' written final projects, and ethnographic field-notes collected by both the instructor and students.[9] The diagnostic test is a mix of thirty-five free-response and multiple choice questions drawn from the Conceptual Survey of Electricity and Magnetism (Maloney et al 2001), and the Electric Circuit Conceptual Assessment (Sokoloff 1999). In addition to selecting answers for each question, students provide confidence levels for their answers on a three point Likert-like scale (guessing, somewhat sure, certain). All students (N=13) participated in all forms of evaluation, with the exception of days when students were absent from class.

## II. PARTICIPANTS AND SETTING: THE IDIOCULTURE:

Physics 180, the class on teaching and learning physics is usually limited to approximately 10 students, and generally populated by advanced (junior and senior level) undergraduate physics majors and first and second year graduate students in physics. The course is an upper division elective in the sequence for physics majors, satisfies one of the prerequisites for the teacher education certification program, and can count for graduate level course credit. The only prerequisite for the course is introductory level physics, though many of the students report having taken more classes in electricity and magnetism (E&M). The ten week course is comprised of three curricular components: physics content, theories of teaching and learning, and practical teaching experience.

Each of the three curricular components of the course represents roughly one third of the course. One of the two weekly class sessions focuses predominantly on the study of traditional physics content. The physics content reviews approximately two-thirds of an introductory course in E&M (using texts such as Halliday, Resnick and Walker 1997). The other weekly class session centres on discussion of readings, which fall into several categories: empirical research on learning (McDermott and Schaffer1992), theoretical underpinnings of learning physics (diSessa 1998), or cognitive science of the teaching and learning processes more generally (J.S. Brown et al. 1989). At least once per week, students engage in the laboratory portion of the course: teaching at the pre-college level. The teaching experiences provide students with practical experience both in physics and in theories of teaching and learning. As much as possible, each component is integrated with the others. The lines between the constituent elements of the class are purposely blurred. A student reading about theoretical difficulties in understanding the concept of electric field is encouraged to wrestle with his own understanding of the topic. Furthermore, as much as possible, there is a temporal alignment of the class components. The same week that students study electric fields, they read about traditional student difficulties in understanding the concept of fields, and also attempt to teach the concept to others.

---

[9] Instructor notes documented the in-class activities; whereas, student notes documented their experiences teaching in pre-college environments.

It is worth highlighting some of the norms within this idioculture. Students were heavily encouraged to reflect upon their own thinking, both in order to facilitate their own learning, and to make overt some of the tacit assumptions many make in approaching physics problems. Collaborative work was emphasized, along with group discussion. Explicit effort was spent on constructing a culture where it was safe not to know answers, or to question one's own understanding. The environment supported challenging one's own assumptions and the broader norms of physics (e.g., why most introductory physics textbooks are ordered in the same fashion). The participants in this community were ultimately developing theories of what it means to know physics along with mechanisms for knowing the material. Lastly, a purposeful blurring of the lines among the various roles that student take (learner, teacher, problem solver, etc.) was strongly encouraged. For instance, during the course of a single week students would solve problems in a given content area, reflect on their own understanding of that topic, read about student difficulties with that topic and teach the topic to more novice students.

## II. PROCEDURE: VARYING SITUATIONS AND TASKS:

By varying and studying student learning situations in theoretically motivated ways, we may describe the influence of context on student learning. A variety of situations and tasks were constructed within this idioculture. These included but were not limited to situations that supported: students talking, students testing, students building objects, students developing laboratory materials or homework problems, an instructor lecturing, and students gathering informally to discuss class or teaching responsibilities. The sorts of typical situations that were constructed included:

Seminars.  The gathering of individuals at a seminar table, and the instructor or a student leading a discussion and asking questions about and implications of the material.

After-school Environments.  A variety of tasks occur in the broad setting, or situation, of an after-school high school program, where five university students gather in a local Boys and Girls Club with a variety of supporting resources including light bulbs and batteries, and work sheets such as those developed in *Physics by Inquiry* (McDermott 1996).

Classroom / Laboratory.  At the university level, a related situation encompasses students assigned to the task of teaching each other a given concept . The situation includes other students enrolled in this course critiquing the presentation, or other university students, who have taken no formal physics courses, serving as the subjects of such presentations.

Tasks are heavily influenced by their surrounding situations.  Two important tasks related to the above situations upon which are addressed here are:

Presentation of readings: The task of writing-up questions or points on the readings, or presenting a summary of a given reading to the class.

Teaching: In after-school situations, one of the common tasks requires students enrolled in the university course to teach pre-college students basic physics concepts, such as current conservation using batteries, wires, and light bulbs.  In classroom situations, students teach each other, or 'naïve' students about the concepts such as flux, using a white board and textbooks.

## RESULTS:

### I. IDIOCULTURE:

Students generally did not enrol in this course to remediate their understanding of physics.  All students in the course had passed one, two, or in some cases, three classes in electricity and magnetism.  Nonetheless, all students demonstrated improved understanding of electricity and magnetism. Results of the pre- and post-tests are shown in Figure 4.  The independent axis of the plot lists individual students.  The left-most student, A, had never formally studied the material; the right most student, N, is a fifth year graduate student in physics.  The dashed line indicates a division between physics majors and non-majors.  The dependent axis plots student performance.  The mean pre- and post-test scores are, respectively, 54% ($\sigma$= 25%) and 74% ($\sigma$ = 24%).  The average of individual student gains is 51% ($\sigma$=30%; N=13; p<0.001).

[Figure 4 about here]

Notably, all students improved in their mastery of the basic concepts of E/M, and no student arrived at this course with a comprehensive understanding of foundational concepts of the domain (despite having done well enough to pass one, two, or three courses on the subject).  By comparison, in a study of more than 5,000 college and university students, Maloney et. al (2001) document average course gains of approximately 32% on a similar instrument over the course of an introductory class.  Of course, the classes examined by Maloney et al. were different idiocultures, with different norms, expectations and approaches.  The introductory courses are designed to work with different students and take a far more traditional approach to covering the course material.

By contrast, the structure of Physics 180 is motivated by the belief that students' conceptual expertise is strongly influenced by students' teaching experiences.  In line with this hypothesis, students report improved ability and interest in teaching. During the first and last weeks of class, students turned in "statements of teaching," where they were asked to write a paragraph or two on their approach to teaching.  One typical example of a student reflection on teaching is as follows:

> Student L: ... there seems to be two ways of going about [getting people to learn]. One school of thought is that repetition is how one learns, and the teacher should focus on the most important ideas and go over them repeatedly.  The other methods is to saturate the students with information... I have no opinion on which method works better… - week 1

Student L: I believe that teaching is less telling and more leading through interactive experiences. It is important for a teacher to know the subject material and be able to convey it clearly, but it is equally important for a teacher to be able to prompt students into learning experiences through which students learn on their own, and in the process own the knowledge themselves … Another important duty of a teacher is to provide an environment for the student that is conducive to learning. This may include ... providing groups of students for interaction and making sure the students are learning and not just memorizing by getting involved in the learning process. - week 10

The class holds a heavily constructivist bent, which seeps into the students' consciousness. A significant effort was made, however, to ensure that students wrestled with the theoretical underpinnings of their convictions and the structure of their own teaching experiences. Some of these theories and tools for understanding the teaching / learning process begin to cycle through public communication in the course as demonstrated by an increased use of technical language from the course readings in student field-notes. For example, Student H writes of pre-college students' failure to grasp a lesson, "This might be a consequence of the fact that they were not forced to confront many of their pre-conceptions, come upon a conflict, and resolve it." These sentiments parallel Posner's comments on developing a theory of accommodation (Posner et al. 1982). The student field-note continues, "knowledge ... never really became integrated as a system," which, in this context, appears to refer to diSessa's (1988) notion of knowledge in pieces and Reif's (1986) discussion of knowledge structures. Students adopt strategies from the readings and reflect on their own success and failure to implement these strategies in the teaching environment.

A dynamic set of rules and norms comprise physics 180. These rules are not simply dictated and static, but rather become adopted by the students and are formed as much by the students as the broader context (the physics department and university) and the instructor. Furthermore, each of the members of this idioculture moves through a variety of idiocultures as part of their education and as part of this particular class. Participating in Physics 180 also means teaching in other environments which belong to different idiocultures (such as the after-school and pre-college educational environments).

On mid and end-of-course evaluations, students reflect upon the unusual course norms:

- Active reflection, discussion and externalization of ideas

    'This is really the first class where I have really had to talk about what I think'

    'I've learned enough to revamp my whole style'

- Students are active agents in the design of their educational experience (and learning).

    'The best part of the class is the freedom to design my own fieldwork and [final] project'

    'I appreciated the instructor's willingness to let me explore some of my own ideas and interests'

- The class is not merely about content mastery, but also about how to think about the content, students' own learning, and the education process.

    'This (teaching physics) is what I do! I have a better perspective of my students now... My perspective has changed. When I first started in this course, I laughed at the word 'preconceptions' ....'

As will become evident below, sorting out one's own reasoning is far more valued than providing correct answers. In many instances, students openly report not knowing or understanding basic material (which they have seen and which they have 'mastered' according to prior exam performance). This recognition and its externalization become part of the learning process.

## II. SITUATIONS / TASKS

Because the tasks and situations are so heavily intertwined (since they co-construct each other), they are combined in the present analysis.

### A. TASK I: PREPARATION

Before a class where the principle of capacitance is to be discussed, students are given a

homework assignment:

*Design a lesson plan to teach what you consider to be the key components of capacitance.*

This assignment is accompanied by guidelines for the lesson plan (identify the audience, scope of material to be covered, assumed knowledge of the students coming into the lesson, outcome objectives, etc.). This particular task is set in a broader situation of learning about capacitance. The assignment is offered during the seventh week of the ten week class. Students have resources of the introductory text books, expect to present their findings to the rest of the class, and are encouraged to discuss their approach to the task and their understanding of the material with each other and the instructor beforehand. In general, the idioculture of the class is one that supports active reflection, where each prior homework set has emphasized reflection on both the process and purpose of solving a given problem.

During the next meeting of the class the discussion is captured on audio-tape:

*Instructor:* Why bother with capacitance? Why do we learn about it?

*Student K:*  I had a really difficult time trying to figure out... because I looked at, I looked at, several textbooks and it's like: This is capacitance. It's you know the ratio of charge over potential and then they just rush into throwing all these other equations how you get defined capacitance in a cylinder and a sphere and they never really got to, 'why do we care?' and then they started throwing out 'oh this is how you add it up in series and in parallel,' and I was like, 'we haven't really learned about circuits yet. How are you ...' I was very confused. And... I thought I understood why capacitance was important but after reading [for this assignment] I have no idea.

Immediately following the class, reflections are written by the instructor:

This was not the only instance in the class where students felt like they knew something and despite difficulties in describing it, teaching it, [they] still felt they understood.  K had a great example supporting this […] when we were discussing the homework overtly.  K said that she understood the material until she went to do this assignment and now she doesn't understand it at all.  I asked if she had really understood the material, or only thought she had understood the material.  She smirked and I followed with a pointed question, 'what about writing a lesson plan could have made you forget what you already knew?'

**Analysis**

Here student K has become aware of her own level of understanding through the task presented and the ensuing class discussion.  Furthermore, in this task / situation, she has come to question what constitutes physics knowledge.  No longer are definitions and equations sufficient to understand capacitance, but understanding why capacitance is important and possibly its relation to other concepts are critical features of knowing about capacitance itself.  The task was not a traditional formation, such as to define the parameters of capacitance, or to calculate the capacitance of some object, but rather to develop a lesson plan for others.  In a traditional task formation, such as those just described, K's understanding may well have been sufficient.  To know capacitance in the context of solving a parallel plate capacitance problem may be to know $C = \varepsilon A/d$.  However, in the very different context (situation), developing a lesson plan with the real or imagined prospect of teaching to others, a much more sophisticated understanding of capacitance is required– to know capacitance in this situation means something different.  Knowing capacitance is intertwined with the context in which it is embedded.  Furthermore, the context itself arises from the relations of the fibres (actions / tasks) that K is weaving together- reading the textbook, preparing a lesson plan, discussing the idea in a classroom, and reflecting upon her own thinking and challenging the ideas presented in the text.  It may be seen that from the construction of this context, student K's thinking about the material has become well enough aligned with the concept of capacitance that she may now observe a fundamental rift between her own understanding and a more robust understanding of

capacitance – the purpose of learning capacitance, the 'why do we care?' is absent.

## B. TASK II: PRESENTATION

Later in the same class, Student K is at the whiteboard in front of the class deriving the equation which describes equivalent capacitance for capacitors in series:

$$\frac{1}{C_{eq}} = \sum_i \frac{1}{C_i}$$

The student was following another who had just worked the derivation for equivalent capacitance in a parallel circuit. Students were encouraged to help one another, and given guiding questions by the instructor. They had not prepared specifically for these derivations and were discouraged from using the textbook as reference.

Again the class discussion from audio-tape:

> *Instructor:* And this is a theme that will come up – that you can use voltage up as you said K, which is interesting, although I prefer to have voltage drops ...[in audible] umm and you can't use charge up. So this would be a good way to foreshadow some difficulties that students classically have in circuits too. Anybody have difficulties with this?

> Class: [laughter].. yeah

> *K:* Oh, like just getting to this point I think ... I had never ... I've always ...you know you always just see the end result in the book. And I think going through this [presenting the derivation at the board and discussion] was really helpful for me.

Commentary / Reflection written by the instructor immediately following class:

> K made the comment that it was really important for her to work through the solution on the board. She had always seen the results but never worked the derivation. She had never understood what caps [capacitors] represent or how they interact with each other.

**Analysis**

This example identifies a situation in which student K appears to construct an understanding of capacitors adding in series. Both the task of working a derivation at the board and the class discussion seem to be critical to K's conceptual shift. Again, a weaving together of the tasks forms the situation in which K demonstrates an understanding both of capacitance adding in series (which is shown in results at the board) and of her own level of understanding (which appears in class discussion). Furthermore, K's discussion reveals a number of situations that have failed to impart an understanding of capacitors adding in series – her two (or more) classes in E&M prior to this course. Though Student K alludes to the fact that the derivation is not in the standard textbook, this information (derivation of capacitance in series or parallel) is contained in the majority of introductory text books (both of those used in this course) and in most introductory lecture courses (which are prerequisites for this class). Classical educational conditions seem to fail to impart a long-lasting command of the material. Additionally, this situation provides insight into K's construction of how classical courses are taught (whether or not they are, in fact, taught that way). Presentation of the material required a realignment of her understanding of capacitance with the formal and abstracted concept. In so doing, K's own understanding shifted.[10]

## C. Task III: reading

The final task analyzed was set in a class discussion of a paper on the use of analogies to learn about electric circuits (Gentner and Gentner 1983). Students spent time discussing the utility of various analogies, and Student F was in charge of summarizing of the paper for the class. Following the brief summary, students debated the utility of a water pump/reservoir analogy versus a gravitational potential analogy to represent the effect of adding batteries in series versus parallel.

Class discussion transcribed from audio-tape:

---

[10]     Note that in figure 4, Student K posts gains of 63%.

*Student F:*     Can we just talk about .. like if you have the uhhh two batteries in series.  You get twice the current

*Instructor:*     right

*Student F:*     Which actually [pause] taught me something.  I always thought the batteries in parallel gave you [inaudible]

[class discussion of the water analogy and two batteries in series produce twice the current of a single battery for a fixed load]

*Student F:*     okay but see, I thought it was the opposite of that.  Because I think I was using the wrong model ...

Commentary / Reflection written by the instructor immediately following class:

During this discussion, Student F made a very interesting revelation, [with] which Student J also identified Total lack of conceptual understanding of series and parallel batteries and bulbs.  Student F made the comment that he had thought batteries worked differently until he had read the article.  Still throughout the class he and Student J made little mistakes about the relative bright[ness], voltage, etc. When asked to think about it in terms of the article and the analogies presented, they would get the right answer.  It required conscious effort and thought.

## Analysis

Through the readings, the presentation of the readings, and class discussion, Student F begins to think about his own thinking and develops an understanding of how voltage adds in series.  He questions what he understands, interrupting the class discussion on analogies to clarify his own ideas about voltage addition.  He moves to a more expert level of understanding, and identifies his own conceptual failure (use of a bad model).  In the broader situation of the class discussion, F's reflection upon his own thinking helps him shift conceptual vantage points.  It is this situation, the weaving together of particular tasks, readings, presentations, and discussions, which allows Student

F to think about his own thought process, to recognize a faulty model and to work on re-conceptualizing voltage addition. At first his understanding is not aligned with the task – reporting on the paper. This reporting, however, causes Student F to re-think his own understanding. As he develops a more expert-like model for how batteries behave in series, his understanding remains embedded in the context of the paper and the class discussion. He continues to make mistakes, and recognizes them only in the context of (in reference to the particular task of) the paper. His knowledge of voltage addition is tied to the context. Ultimately, the hope is that Student F transfers his new-found understanding to other situations appropriately. He appears to do so in final course assessment.[11]

## DISCUSSION

As found with other researchers (e.g., diSessa et al. or Leander and Brown), these data suggest that learning is critically grounded in context. Conceptual change is a more complicated process than can be captured by models which purely focus on student- and/or content-centred approaches such as the theory of accommodation of new ideas as realized by an elicit-confront-resolve model (Posner et al. 1982). In the first two situations above, student K knows something in one context (working problems in a textbook or calculating equivalent capacitance); however when the situation changes, where K becomes responsible for presenting these concepts, she recognizes that she no longer knows the material. In these cases both student learning and the concept (capacitance) change with context. Context, the situation of presenting to others, co-constitutes K's understanding and the concept itself. From the more traditional theory of accommodation, context does not play a role. Students' knowledge is bimodal; they understand or do not, irrespective of context. Such a theory is insufficient for describing K's learning about capacitance. In the third situation examined, it may be argued that Student F might never have learned the rules for series and parallel addition of voltage sources. Furthermore, perhaps student F simply needed the appropriate cognitive conflict to spur him to develop a more robust theory (via accommodation). Whether or not this is the case, it is clear that Student F's conceptual change (and *all* conceptual change) occurs within a given context. His interaction with the material and the others in this situation cause him to re-develop a model for voltage. Importantly, this conceptual shift occurs within a given context – reading and discussing a paper about student difficulties with electric circuits.

---

[11] In the conceptual survey of the course material, Student F improves 55% and in particular, on series and parallel circuit questions Student F improves 33% (from 70% to 80% correct). Furthermore his confidence level increases from 1.8 to 1.1, where 1 is certain and 3 is guessing.

What allows for, promotes, and is created by these conceptual shifts is the aggregation and relations of such situations -- an idioculture with particular norms, values, and practices. In the idioculture of Physics 180, not knowing is acceptable, questioning understanding is valued, and student engagement is the rule. These norms allow for students to engage in situations which support conceptual change and in turn the collection of these situations over time help form the idioculture. Ultimately, a primary educational goal is for students to abstract from this local context (situations and idioculture) and apply (and re-contextualize) this knowledge in another – what others refer to as transfer.

If one attempts to ascertain *why* students have particular models, or wishes to create environments which promote conceptual change in students, learning must be considered *in situ*, with context as part and parcel of student learning. Learning is not an isolated action, but rather a social activity influenced by the local context, task , situation, and idioculture. Inherent in a given context are certain constraints which promote or inhibit construction of content understanding. From this perspective, conceptual change is mutually constituted by the individual and the context. An important goal of this research is to create environments where the context supports student learning of concepts — that is, to create an environment with sufficient constraints and opportunities for a student engaged in an activity that the student becomes closely enough aligned with the concept at-hand. This sufficient alignment assures that the concept itself becomes instrumental and motivated for the student. The concept becomes a tool for the task at-hand. Connected with the concept are the symbols and abstractions (e.g., equations or diagrammatic representations) necessary to allow the student to apply this concept to various other situations in useful manner – i.e. which allow transfer. In short, the environment should support both the development of conceptual understanding in a particular situation and the ability to transfer such understanding to relevant new situations.

By attending to context explicitly, it becomes possible to identify situations that are particularly effective, why they are effective, and what it means to replicate these reforms. Both here and within other efforts in physics education reform, situations (such as teaching have been found to be successful mechanisms for supporting student learning. As researchers pursue the design of learning environments which support such learning situations, two lines of research emerge. The first line of research is to examine why these situations are effective. This line of inquiry would seek to build on models of student cognition and incorporate the above described notions of context --- For instance following the lines of diSessa et al, how does student epistemology vary by task, situation and idioculture? Or, building on Leander and Brown, how might (in)stabilities be understood in terms of the dynamics of tasks, situations and idiocultures?

The second line of inquiry would be to foreground context in study of sustaining and scaling successful educational reforms in physics. How might these evocative situations be incorporated in a large-scale, practical manner? Or, given a successful educational intervention, which features and dynamics (contexts) are necessary to replicate reforms? An understanding of context can provide insight as why so many attempts to sustain and replicate successful educational programs fail. The developed model of context presents some insights as to why educational reforms can neither be stamped out, nor made teacher-proof – the reforms themselves, whether at the level of situational activities, or broader idiocultures, exist within and shape broader contexts. What works in one university setting does not necessarily work at another. The students and resources are different, and the institutional structures surrounding the reforms are different. Hence educational reforms must be dynamic, nesting among local contexts (or boundary conditions) and remain flexible enough to evolve with the local cultures in which they are being enacted. Attending to such contextual considerations would potentially provide a mechanism by which researchers and reformers in higher education might share successful approaches to teaching physics.

## CONCLUSIONS

This paper includes context as an integral factor for understanding learning physics in the university physics classroom; it is not appropriate to discuss student learning absent from the context with which it is intertwined. I am not claiming that the idea of context affecting student learning is new, but rather this paper develops a theoretical framework to describe what is meant by context as it pertains to students' learning in a university level physics classroom. Many reform efforts have acknowledged implicitly the role of context, local situation and even local culture. Workshop Physics, Physics by Inquiry, Constructing Physics Understanding, Modelling Workshops and a host of other reforms documented by McDermott and Redish are successful in part because they construct contexts which support student learning (McDermott and Redish 1999). This paper, and others that examine student learning in context (Otero 2004), begins to develop some tools that will hopefully allow physics educators to describe not only what is happening in their classrooms, but why and how it is happening. While such tools are applicable to many different theoretical frameworks for student learning, their essential contribution is re-contextualizing the discussion to include the inextricable role of context.

## ACKNOWLEDGMENTS:

This research was conducted with the support of the National Science Foundation's Post-doctoral Fellowships in Mathematics, Science, Engineering, and Technology Education (NSF's PFSMETE Grant Number: DGE-9809496) under the mentorship of Michael Cole (University of California, San Diego) and Andrea diSessa (University of California, Berkeley). I wish to thank these mentors, my colleagues at the Laboratory of Comparative Human Cognition UCSD and the Physics Education Research Group at Colorado (CU Boulder) for intriguing discussions, insights, and their support. I wish to additionally thank the anonymous reviewers for their thoughtful, extensive and constructive comments on earlier drafts of this work.

**Figure captions:**

Figure 1: Embedded levels of context.  A student engaged in a task is embedded in the context of a broader activity, which is part of a class, which is embedded in a university, etc.

Figure 2: Three frames of context: task, situation, and idioculture

Figure 3: Idioculture as both surrounding and constituted by situations

Figure 4: Student achievement on conceptual assessment of electricity and magnetism.

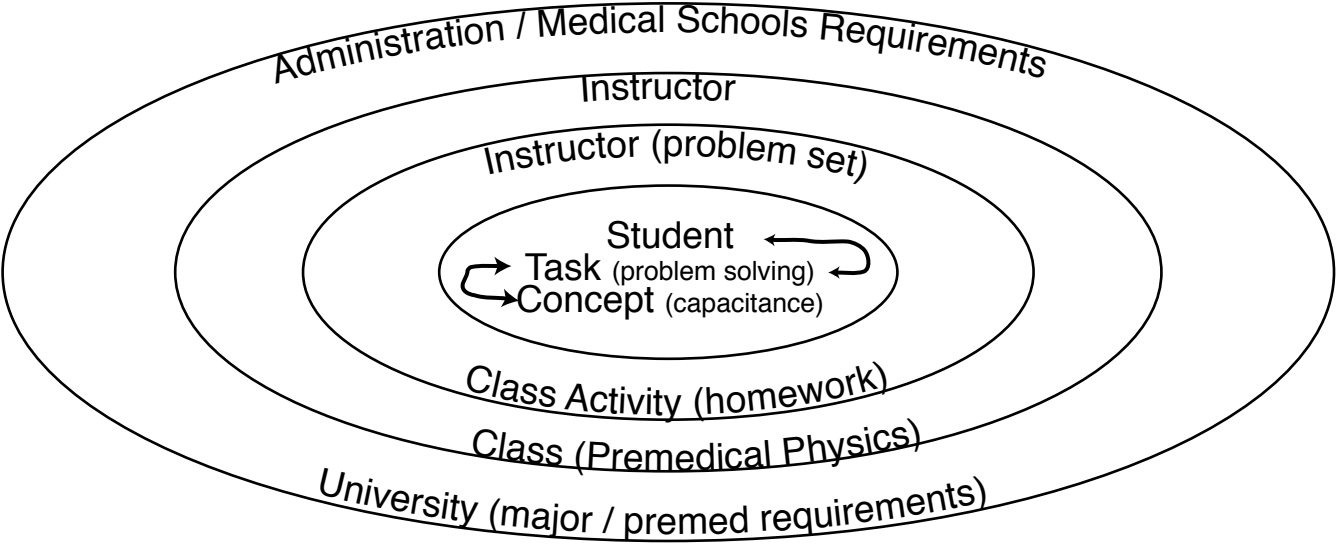

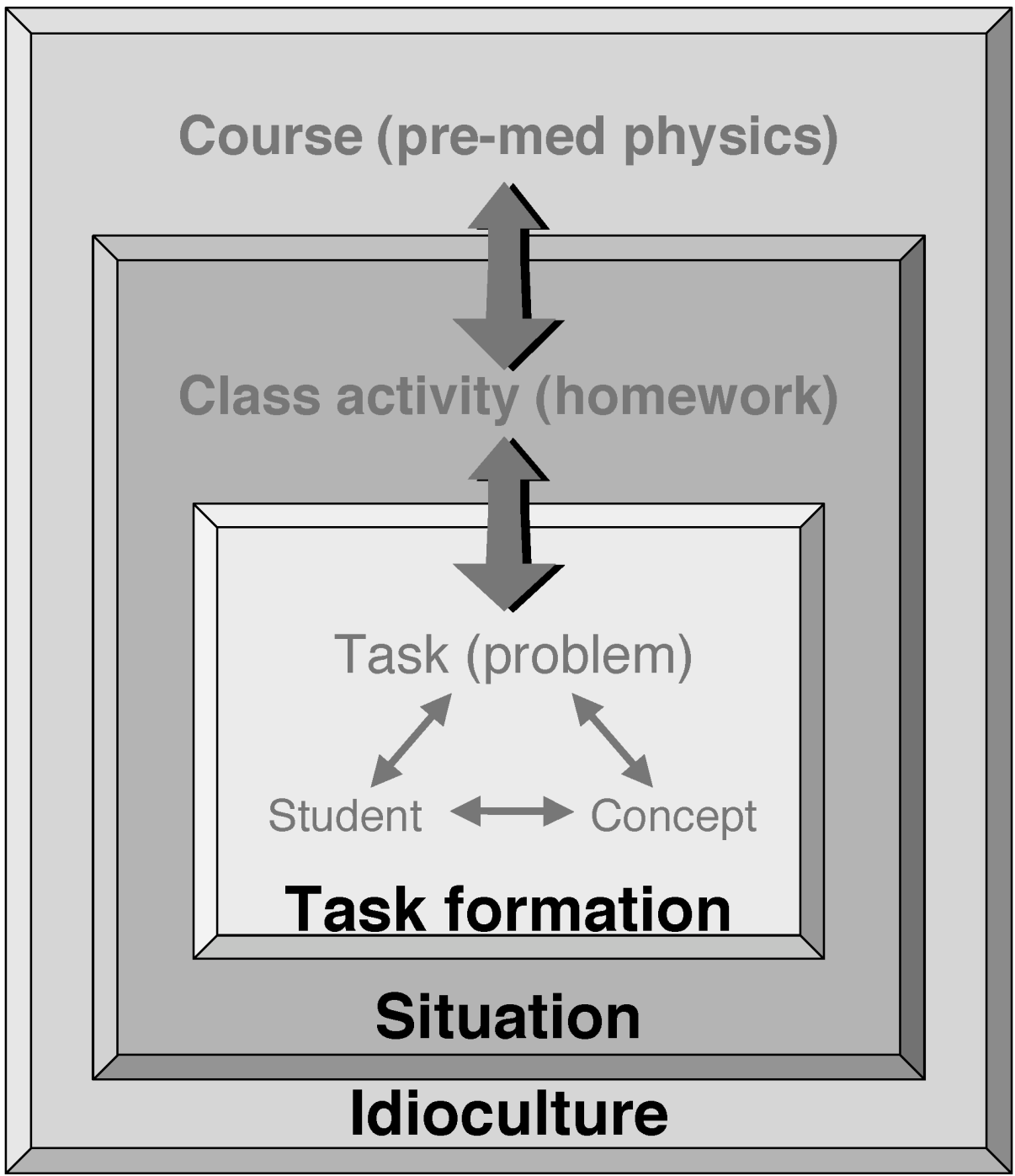

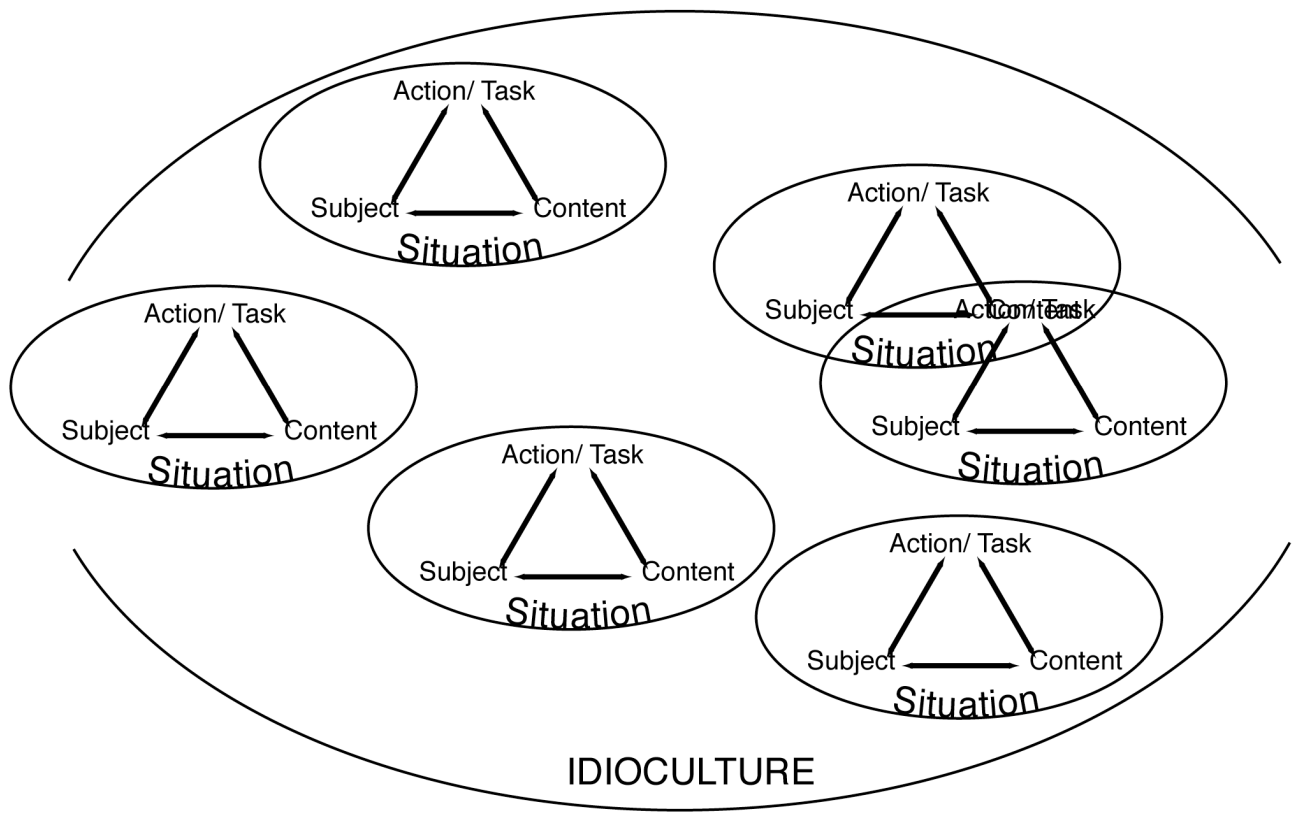

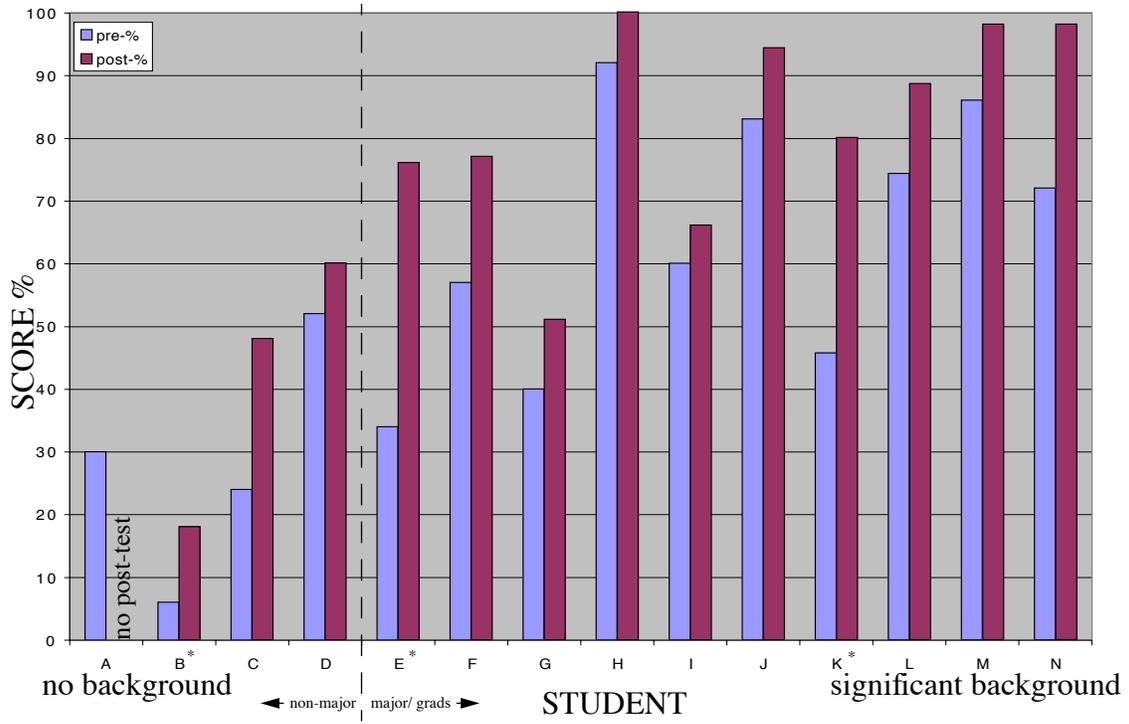